\documentclass[12pt]{iopart}
\usepackage{graphicx}
\usepackage[pdftitle={Magnetic Dipole Transitions with the Full Skyrme Interaction},pdfauthor={M. C. Barton},colorlinks=true,citecolor=black,linkcolor=black]{hyperref}
\usepackage{color} 
\usepackage{float}
\usepackage{pbox}
\usepackage{braket}
\begin{document}
\title[Magnetic dipole transitions using the full Skyrme interaction]{Magnetic dipole transition calculations in $^{52}$Cr using the full Skyrme interaction}
\author{M.~C.~Barton and P.~D.~Stevenson}
\address{Department of Physics, University of Surrey, Guildford, Surrey, GU2 7XH, United Kingdom}
\date{\today}
\begin{abstract}
The purpose of this paper is to investigate the use of the TDHF method with different Skryme force interactions for the prediction of the M1 strength function associated with $^{52}$Cr, in particular due to the tensor part of the Skyrme force.$^{52}$Cr was chosen due to the recent interest experimentally in this nucleus and the clear data avaliable for comparison.

The method used involves applying an instantaneous boost to the wavefunction at the beginning of the dynamical calculations in order to initiate a broadband excitation of the nucleus, following which a Fourier analysis is made.

The results of the TDHF simulation found broad agreement with experiment for the range of energies which excite an M1 transition for a variety of forces. The tensor terms appear to be important in qualitatively changing the details of the strength distribution in the main part of the strength function between around 8 and 12 MeV.
\end{abstract}
\section{\label{sec:intro}Introduction}
\noindent The time-dependent Hartree-Fock (TDHF) technique \cite{Sim12} using Skyrme forces is a versatile method for investigating many collective nuclear phenomena, such as heavy ion collisions \cite{UO10a,SY13,Uma06,coll} and giant resonances \cite{Ste04,Ste11,elecdip}. It is an investigation of the latter, using TDHF, with which this paper is concerned.  In particular, the focus is on giant magnetic dipole resonances. The main purpose is to investigate the effect that the tensor force present in the Skyrme interaction has on the prediction and reproduction of these resonances.  Giant resonances in general, and magnetic dipole resonances in particular are often studied with methods derived from the Random Phase Approximation \cite{srpa} or the Shell Model \cite{reviewmag}, with time-dependent methods giving a complementary picture. For the present purposes, $^{52}$Cr (Z=24, N=28) has been chosen to allow easy comparison with experimental data \cite{CR52,moderncr52}, as well as between assumptions in the form of the effective interaction, while being a spin-orbit unsaturated nucleus in which effects of the tensor interaction would not be masked.

\section{\label{sec:exp}Observed Strength} For the comparison of the presented calculations with data, two experiments are drawn upon: An older electron scattering experiment (which we refer to as ES) \cite{CR52} and a newer photon scattering experiment (referred to as PS) \cite{moderncr52}.  The ES data span excitations energies in $^{52}$Cr between 7 and 12 MeV, while the PS data covers an excitation range 5 -- 9.5 MeV.   In the ES case, the multipolarity of resonances was not always unambigusous and we take those transitions assigned categories \textit{A} and \textit{B} in \cite{CR52} as being of wholly or predominantly $M1$ in character.

\section{Theory}
\subsection{Skyrme Energy Density Functional}
\noindent The calculations in this paper used the phenomenological Skyrme force. The two-body non-tensor and tensor parts of the full Skyrme force are given, respectively, by \cite{skyrme}
\begin{eqnarray}
\upsilon_{12}^{(2)} \ &=&\ t_{0} (1+x_{0}P_{\sigma}) \delta ( \mathbf{r}_{1} - \mathbf{r}_{2}  ) \\ &&+ \ \frac{t_{1}}{2} \Big( 1 + x_{1} P_{\mathbf{\sigma}} \Big) \Big[\delta ( \mathbf{r}_{1} - \mathbf{r}_{2} ) \mathbf{k}^{2} + \mathbf{k}'^{2} \delta ( \mathbf{r}_{1} - \mathbf{r}_{2} ) \Big] \\&& + \ t_{2} \Big( 1 + x_{2} P_{\mathbf{\sigma}} \Big) \mathbf{k}' \cdot \delta ( \mathbf{r}_{1} - \mathbf{r}_{2} ) \mathbf{k} \\ &&+ \ i W_{0}(\mathbf{\sigma}_{1} + \mathbf{\sigma}_{2}) \cdot \Big( \mathbf{k}' \times \delta ( \mathbf{r}_{1} - \mathbf{r}_{2} ) \mathbf{k} \Big) + \\
&&\frac{t_{3}}{6} (1 + x_{3} P_{\sigma} ) \rho^{\alpha} (\frac{\mathbf{r}_{1} + \mathbf{r}_{2}}{2} ) \delta (\mathbf{r}_{1} - \mathbf{r}_{2})
\end{eqnarray}
\begin{equation}
\eqalign{
v_{t} \ = \ \frac{t_{e}}{2} \bigg\{ \big[ 3 (\mathbf{\sigma}_{1} \cdot \mathbf{k}')(\mathbf{\sigma}_{2} \cdot \mathbf{k}') - (\mathbf{\sigma}_{1} \cdot \mathbf{\sigma}_{2}) \mathbf{k}'^{2} \big]\delta ( \mathbf{r}_{1} - \mathbf{r}_{2}  ) \\ + \big[ 3 (\mathbf{\sigma}_{1} \cdot \mathbf{k})(\mathbf{\sigma}_{2} \cdot \mathbf{k}) - (\mathbf{\sigma}_{1} \cdot \mathbf{\sigma}_{2}) \mathbf{k}^{2} \big]\delta ( \mathbf{r}_{1} - \mathbf{r}_{2} ) \bigg\} \\ 
+ t_{O} \bigg\{ \big[ 3 (\mathbf{\sigma}_{1} \cdot \mathbf{k}')(\mathbf{\sigma}_{2} \cdot \mathbf{k}) - (\mathbf{\sigma}_{1} \cdot \mathbf{\sigma}_{2}) \mathbf{k}' \cdot \mathbf{k} \big] \delta ( \mathbf{r}_{1} - \mathbf{r}_{2}  ) \bigg\} \\}
\end{equation}
where 
\begin{equation}
\mathbf{k} \ = \ \frac{1}{2i} (\mathbf{\nabla}_{1} - \mathbf{\nabla}_{2}) \quad{} \quad{} \mathrm{and} \quad{} \quad{} \mathbf{k}' \ = \ - \frac{1}{2i} (\mathbf{\nabla}_{1}' - \mathbf{\nabla}_{2}')
\end{equation}
\begin{equation}
\bra{\phi} \mathbf{k}' \ket{\phi} \ = \ \int d^{3} r \phi^{*} \mathbf{k}' \phi \ = \ - \frac{1}{2i}  \int d^{3} r \Big[(\mathbf{\nabla}_{1} - \mathbf{\nabla}_{2})\phi^{*} \Big]  \phi
\end{equation}
with $\mathbf{k}$ acting to the right and $\mathbf{k}'$ to the left.

There are many parameterisations of this force, fitted by different algorithms to varied sets of data and pseudodata.  In the present work, SkM* \cite{skms}, Sly5 \cite{sly5}, Sly5t \cite{sly5t},T12 \cite{Tpar} and T22 \cite{Tpar} are used in order to assess a widley-used force in which the tensor terms are inactive (SkM*), to compare a force which was originally fitted without the tensor terms (SLy5) to which tensor terms were later added without refitting the rest of the force (SLy5t) and forces which were fitted including the tensor terms (T12 and T14).

Using the Skyrme force, one can construct an energy density functional which, through a variational principle, can in turn be related to a single-particle Hamiltonian, which is then used to solve for the single-particle states. This can be used to create an energy density functional of many gauge invariant quantities. In particular, we consider the effects of the following two gauge-invariant terms in the energy density functional:
\begin{equation}
\eqalign{
\frac{t_1x_1+t_2x_2-2(t_e + t_o)}{8} \Big( \mathbf{s}(r) \cdot \mathbf{T} (r) -  \sum_{\mu \nu} J_{\mu \nu} (r)^{2} \ \Big) + \\
\frac{t_2 - t_1 +2(t_e - t_o)}{8}  \sum_{q} \Big( \mathbf{s}_{q} (r) \cdot \mathbf{T}_{q} (r) - \sum_{\mu \nu} J_{q, \mu \nu} (r)^{2} \Big) 
}
\end{equation}

and

\begin{equation}
\eqalign{
\frac{3(t_{e} + t_{o})}{4}  \Big( \mathbf{s} (r) \cdot \mathbf{F} (r) - \frac{1}{2} \sum_{\mu \nu} J_{\mu \nu} (r) J_{\nu \mu} (r) - \frac{1}{2} \sum_{\mu} J_{\mu \mu} (r)^{2} \ \Big) \\
- \frac{3(t_{e} - t_{o})}{4}  \sum_{q} \Big( \mathbf{s}_{q} (r) \cdot \mathbf{F}_{q} (r) - \frac{1}{2} \sum_{\mu \nu} J_{q, \mu \nu} (r) J_{q, \nu \mu} (r) - \frac{1}{2} \sum_{q \mu } J_{q, \mu \mu} (r)^{2} \Big),}
\end{equation}
which we refer to as the $\mathbf{s}\cdot\mathbf{T}$ and $\mathbf{s}\cdot\mathbf{F}$ terms respectively.

A more thorough description the Skyrme energy density functional including the definitions of the densities given above and corresponding derivation of gauge invariant quantities is given in \cite{Bender}.
\subsection{Hartree-Fock}
\noindent In order to obtain the ground states from which to begin the time-dependent calcualtions, the {\textit Sky3D} code \cite{sky3d} is used in static mode, with wave functions initialised as harmonic oscillator states, and iterated until self-consistent convergence.
\subsection{Magnetic Dipole Resonances}
\noindent The form of the magnetic dipole (M1) operator that has been used for this work is given by \cite{M1operator}
\begin{equation}
\eqalign{
\hat{M}_{10} \ = \ \mu_{N} \sum_{i=1}^{A} \Big\{ g_{s}^{(i)} \hat{s}_{i} + g_{l}^{(i)} \hat{l}_{i} \Big\} \cdot \Big( \nabla r Y_{10} (x,y,z) \Big)_{r=r_{i}} \\
Y_{10} (x,y,z) \ = \ \frac{z}{r} \\ 
\hat{M}_{10} \ = \ \mu_{N} \sum_{i=1}^{A} \Big\{ g_{s}^{(i)} \hat{s}_{i} + g_{l}^{(i)} \hat{l}_{i} \Big\} \cdot \Big( \hat{z} \Big)_{r=r_{i}} \\
\hat{M}_{10} \ = \ \mu_{N} \sum_{i=1}^{A} \Big\{ g_{s}^{(i)} \hat{s}_{z,i} + g_{l}^{(i)} \hat{l}_{z,i} \Big\} \\
g_{s}^{(i)}  = g_{p}, \ g_{l}^{(i)} = 1 \quad{} \textrm{for protons} \\ \textrm{and} \\ 
g_{s}^{(i)} = g_{n}, \ g_{l}^{(i)}  = 0 \quad{} \textrm{for neutrons} \\
\hat{l}_{z,i} = \frac{1}{i} \Big( x \frac{d}{dy} - y \frac{d}{dx} \Big)}
\end{equation}
$g_{p}$ and $g_{n}$ are the g factors for the proton and neutron respectively, $\mu_{N}$ is the nuclear magneton, $s_{z,i}$ and $l_{z,i}$ are the spin and orbital $z$ components respectively and $Y_{10}$ is a spherical harmonic. To see the effect of a magnetic transition, all frequencies of this transition can be excited at once using the following boost to the wavefunction
\begin{equation}
\eqalign{
\psi_{new} (t=0) \ = \ e^{i \eta \hat{M}_{10}} \psi_{old} (t=0).}
\end{equation}
To keep within the linear response regime, a variable $\eta$ is added to control the strength of the boost to the wavefunction. The expectation value of the operator $M_{10}$ can be related to the strength function (the strength function gives prediction of how likely a transition is likely to occur for a given frequency or energy) associated with the transition, by \cite{strength} 
\begin{equation} \label{strength}
S ( E ) \ = \ - \frac{1}{\pi \eta} \mathrm{Im} (\mathcal M(\hbar \omega)).  
\end{equation}
in which $ \mathcal M $ is the fourier transform of the expectation value of $M_{10}$ over time. This formula linking the strength function to expectation value in equation \ref{strength} is only valid within the linear regieme.  
\section{Theoretical Results}
\subsection{Hartree-Fock Case Calculations}
\begin{table}[b]
\centering
\begin{tabular}{| l | l | l | l |}
\hline 
Force        & B.E./A [MeV] & Charge Radius [fm] & $\beta_{2}$ \\  \hline
Sly5         & $-8.427$  & $3.708$       &   $0.157$ \\ \hline
Sly5t        & $-8.509$  & $3.700$       &   $0.148$ \\ \hline
SKM*         & $-8.411$  & $3.710$       &   $0.160$ \\ \hline
T12          & $-8.423$  & $3.706$       &   $0.156$ \\ \hline
T22          & $-8.426$  & $3.705$       &   $0.156$ \\ \hline
Experimental & $-8.7762$ & $3.643(3)$    &   $0.211$ \\  \hline
\end{tabular}
\caption{Static (HF) output from Sky3D code, along with experimental results \cite{PhysRevC.27.113,Wapstra1971267,chartofnuclides}.}
\end{table}
\noindent Hartee-Fock calculations were performed on a 3-dimensional Cartesian grid from $-11.5$ fm to $+11.5$ fm, in $1$ fm steps in each direction. The above table outlines the properties of $^{52}$Cr predicted in the static case for various Skyrme forces alongside known experimental data. The results produced are in relative agreement with experiment. The charge radius $r_{rms}$  was approximately calculated from the nuclear proton radius $r_{nucp}$ using the standard formula,
\begin{equation} \label{chargeradius}
r_{rms}^{2} \ = \ <r^{2}>_{nucp} + r_{p}^{2},
\end{equation}
with a value of the proton radius $r_p^2=0.64\mathrm{fm}^2$ adopted from a recent review \cite{RevModPhys.75.121}
\subsection{Sly5t and Sly5}
\begin{figure*} 
\includegraphics[width=\textwidth]{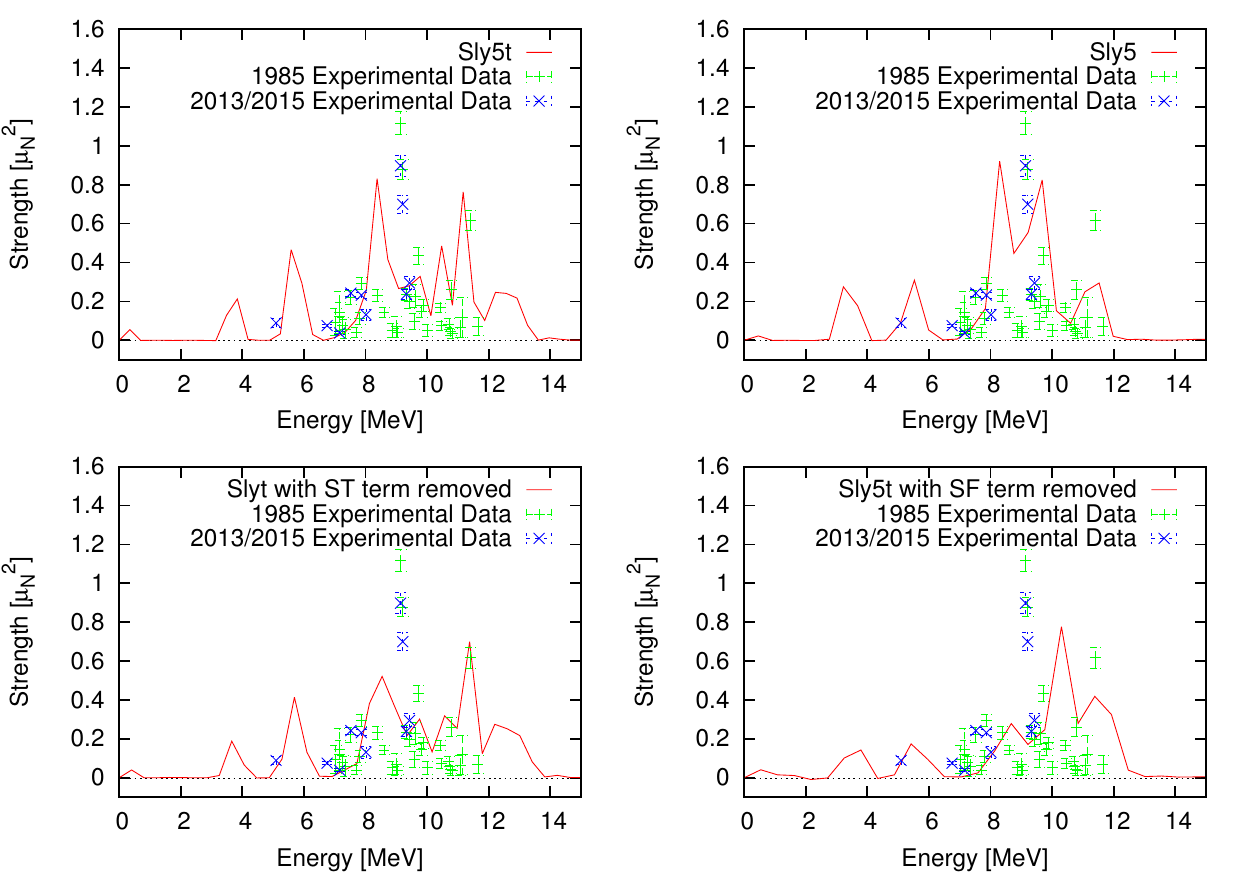}
\caption{The strength functions for the Sly5 and Sly5t parameterisations of the Skyrme force. The strength functions for Sly5t with different Galilean invariant terms removed.\label{sly5variants}}
\end{figure*}

\noindent The tensor force present in the Skyrme force has in the past been disregarded in many calculations as it is assummed to have a small effect compared to other terms in the Skyrme force. Comparing the strength functions for Sly5 and Sly5t in figure \ref{sly5variants}, it can be see that the effect adding the tensor force is to redistribute the strength over a larger energy range. For example the maximum energy where there is any appreciable strength is approximately 12 MeV where as for Sly5t this energy is approximately 13 MeV.  
\subsection{Instabilities}
\noindent While performing the calculations in this paper, an instability was encountered. We suppose that this instability is due to the known problems in the spin orbit part of the Skyrme force \cite{sly52}. The we thus tested the version of the Sly5 which was adjusted to push the instability further away \cite{sly52}. For $^{52}$Cr is was found to be much more stable and allowed a higher resolutions strength function to be produced. It should be noted though that for other nuclei, the instability still existed. The strength function produced by this force are shown in figure \ref{differentforces}. 

\begin{table}[bt]
\centering
\begin{tabular}{| l | l | l | }
\hline 
Force        & \pbox{20cm}{$_{}$\\ Minimum \\ Enegy [MeV] \\ \vspace{1mm} } &  \pbox{20cm}{$_{}$\\ Maximum \\ Energy [MeV] \\  \vspace{1mm}}  \\  \hline
Sly5         & 3.00  & 12.00      \\ \hline
Sly5t        & 3.00  & 13.00       \\ \hline
Sly5t SF OFF & 3.00  & 12.50       \\ \hline
Sly5t ST OFF & 3.00  & 13.00       \\ \hline
SKM*         & 2.50  & 11.50        \\ \hline
T12          & 3.00  & 12.50        \\ \hline
T22          & 3.00  & 12.50       \\ \hline
Experimental & 5.10 & 11.65  \\ \hline
\end{tabular}
\caption{Approximation of minimum and maximum energy of appreciable strength.}
\end{table}
\subsection{Contribution of Different Galilean Invariant 	quantities}
\noindent The particular combination of terms in the Skyrme interaction that give rise to gauge-invariant quantities may each be turned off or on as one chooses.  Since some of these concern only time-odd quantities that are not fitted in the original definition of the forces, they are often set to zero for even for situations when they might otherwise be active, as is the case here.  Therefore the individual effects that these components (defined as the $\mathbf{s} \cdot \mathbf{T}$ and $\mathbf{s} \cdot \mathbf{F}$ terms, above) have on the strength function was investigated. The results with these terms turned on (Sly5t) or off (labelled in figure as ``with ST term removed'' etc) are shown in figure \ref{sly5variants}.  The main result appears to be that the $\mathbf{s} \cdot \mathbf{F}$ term is responsible for a large part of the strength in the region around 8-10 MeV.  There is considerable strength here from the full Sly5t force, which is much diminished by removal of the $\mathbf{s} \cdot \mathbf{F}$ term.
\subsection{Different Skyrme forces}
\begin{figure*} 
\includegraphics[width=\textwidth]{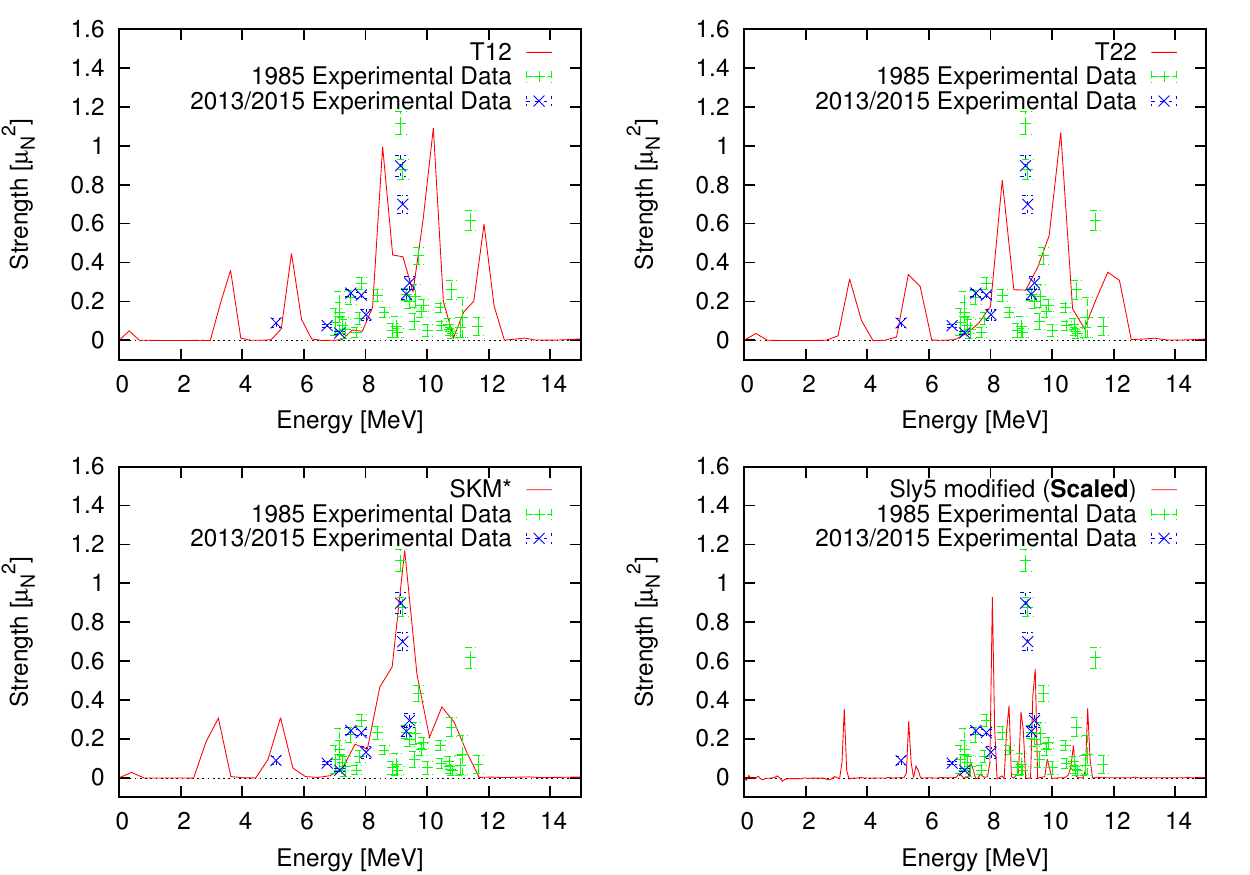}
\caption{The strength functions produced for the parameterisations of the Skryme force T12, T22 and SKM*, along with the modified version of Sly5 force designed to remove instabilities.\label{differentforces}}
\end{figure*}

\noindent It was found that by changing the parameterisation of the force changed the distribution of the strength with the same energy region. The SKM* parameterisation appears to most closely replicate the energy region that has been experimentally been examined. The T12 and T22 parameterisations give similar strength functions. The resultant strength functions are shown in figure \ref{differentforces}. 
\section{Conclusions\label{sec:conclusions}}
\noindent In conclusion it was found that the use of TDHF with Skyrme forces to model magnetic dipole transitions in the linear regime, at least in the case of $^{52}$Cr is useful for predicting the range of energies which may initiate a magnetic dipole transition. An intereasting thing to note is all the Skyrme forces predicting a transition at approximately 3 MeV, which presumably corresponds to a bound state. This study could be further expanded by looking at the recent experimental data for $^{50}$Cr \cite{50Cr} or to expand the code to include higher order multipoles.

\ack
The authors gratefully acknowledge support from the UK Science and Technology Facilities Council (STFC) via awards ST/M503824/1 and ST/L005743/1 as well as computing time on the STFC DiRAC computer.

\end{document}